\documentclass[12pt]{article}
\usepackage{relsize}
\usepackage{mathdots}
\usepackage{amsfonts,amsmath,amssymb}
\usepackage{scalerel}[2016/12/29]
\usepackage{graphicx}
\usepackage{tikz}
\usepackage{float}
\usepackage{url}
\usepackage{braket}

\usetikzlibrary{arrows}
\usetikzlibrary{decorations.markings}

\newcommand{\hb}{\hskip -0.05cm}
\newcommand{\bb}{\hskip -0.1cm}

\newcommand{\hp}{\hskip -0.05cm + \hskip -0.05cm}
\newcommand{\hm}{\hskip -0.05cm - \hskip -0.05cm}

\newcommand{\he}{\hskip -0.05cm = \hskip -0.05cm}

\newcommand{\bi}{\begin{itemize}}
\newcommand{\ei}{\end{itemize}}

\def\be{\begin{equation}}
\def\ee{\end{equation}}
\def\bea{\begin{eqnarray}}
\def\eea{\end{eqnarray}}
\def\tr{{\rm tr}\,}
\def\half{{\textstyle {1\over 2}}}

\def\rme{\mathrm{e}}
\def\rmi{\mathrm{i}}
\usepackage{lineno}
\usepackage{xcolor}
\usepackage{lineno}

\usepackage{amsmath}
\usepackage{epsfig,graphicx}
\usepackage{graphicx}
\usepackage{tikz}

\def\be{\begin{equation}}
\def\ee{\end{equation}}
\def\bea{\begin{eqnarray}}
\def\eea{\end{eqnarray}}
\def\Tr{{\rm Tr}\,}

\def\ui{{u^{-1}}}
\def\vi{{v^{-1}}}

\def\Q{\mathrm{Q}}

\def\rme{\mathrm{e}}
\def\rmi{\mathrm{i}}

\begin{document}

\thispagestyle{plain}

\title{Algebraic area enumeration for open lattice walks}

\author{
St\'ephane Ouvry$^{*}$ 
\and 
Alexios P. Polychronakos$^{\dagger}$
}

\date{\today}

\maketitle

\begin{abstract}
{We calculate the number of open walks of fixed length and algebraic area on a square planar lattice by
an extension of the operator method used for the enumeration of  closed walks. The open walk
area is defined by closing the walks with a straight line across their endpoints and can assume half-integer values
in lattice cell units. We also derive the length and area counting of walks with endpoints on specific
straight lines and outline an approach for dealing with walks with fully fixed endpoints.}
\end{abstract}

\noindent
* LPTMS, CNRS,   Universit\'e Paris-Saclay, 91405 Orsay Cedex, France\\
${ }$\hskip 0.38cm{\it stephane.ouvry@u-psud.fr}

\noindent
$\dagger$ Physics Department, the City College of New York, New York, NY 10031, and\\
${ }$\hskip 0.35cm The Graduate Center of CUNY, New York, NY 10016, USA
\\ \noindent
${ }$\hskip 0.38cm{\it apolychronakos@ccny.cuny.edu}


\section{Introduction}

The enumeration of walks of a given length and enclosed algebraic
area on various lattices is a fascinating problem \cite{bibi}. It is of interest in pure combinatorics, where it has its
own justification in terms of counting of objects among pre-defined ensembles, but also in
physics, as it maps to various models in quantum and statistical physics. The most famous among these
models is the Hofstafdter Hamiltonian \cite{hof}, which describes the dynamics of a quantum particle hopping on a lattice
in a perpendicular magnetic field and leads to the celebrated ``butterfly" energy spectrum.
In addition, the lattice walk enumeration problem is related \cite{nous} to even more exotic quantum concepts
such as exclusion statistics \cite{Hal}, which generalizes standard Fermi statistics to particles obeying
a stronger exclusion principle parametrized by an integer $g \ge 2$ (usual Fermi statistics is $g = 1$;
$g = 2$ is relevant for square lattice walks).

The connection to quantum statistics offers an approach for addressing the enumeration problem of
more general classes of
walks on various lattices. For example, chiral walks on a triangular lattice, with a nonhermitian Hofstadter-like
quantum Hamiltonian generating the biased hopping on the lattice, have been analyzed \cite{nous} in terms of
particles obeying $g = 3$ exclusion statistics. Walks on the honeycomb lattice can also be examined \cite{hexagon} in this
framework and involve particles obeying a mixture of Fermi and $g = 2$ exclusion statistics.
These developments have also born new resuls in quantum statistics, leading to a full description of exclusion
statistics particles in a nondegenerate discrete 1-body quantum spectrum and to explicit
expressions for their cluster coefficients and ensuing thermodynamics \cite{nousbis}.
In addition, this approach establishes a connection \cite{nousdeux} between planar lattice walk enumeration and specific
one-dimensional walks (``paths") of generalized Dyck or Motzkin form (also known as Lukasiewicz paths \cite{theguy}).
The various algebraic enumeration formulae obtained for walks admit a combinatorial interpretation in terms of
paths \cite{nouster}, which leads to cross-pollination between the two subjects.

The algebraic area enumeration was up to now restricted to closed walks that return to their starting point on
the lattice. In that case the algebraic area is the area enclosed by the walk weighted by the winding number
of the walk around each area patch. However, it is also possible to define the algebraic area of an {\it open} walk
using a ``closing'' prescription, the most natural one being joining its two endpoints by a straight line. In the case of square lattice walks,
it is easy to see that the algebraic area defined this way is always an integer or half-integer in units of elementary
lattice cells. 

Once the algebraic area of an open walk is defined, the enumeration of open walks of given length and
algebraic area starting from a fixed point on the lattice becomes an interesting issue. In principle,
the tools developed earlier for the case of closed walks \cite{bibi,nous} are not directly applicable to open walks, and their
enumeration for fixed length and area was an open problem. Still, we will demonstrate in this work
that the problem of open walks on the square lattice can be recast in terms quite similar to the closed walk case
through the introduction of a new operator $\sigma$, in addition to the usual $u, v$ hopping lattice operators
that were useful for closed walk enumeration, thus enlarging the usual ``quantum torus'' algebra $u v =\Q\, v u$
to a ``reflection quantum torus'' algebra. This algebra also appeared, in a different guise and representation,
in the enumeration of {\it closed} walks on the honeycomb lattice \cite{hexagon}, and it is remarkable that it reappears
in the seemingly different context of open square lattice walks.

In the sequel we will calculate the generating function of planar open walks weighted by their algebraic
area and will derive explicit formulae for the multiplicity of open walks of given length and area. 
The case of walks with endpoints fixed on a straight line is also readily tractable with our
method, and we will derive the corresponding generating functions. We will also briefly address he problem of
walks with {\it fully} fixed endpoints and will highlight possible approaches for a complete algebraic area enumeration. 

\section{Open walks}

We consider open random walks of fixed length  $\bf n$ (number of steps) on the square lattice starting
at the origin and ending at an arbitrary lattice point $(k,l)$.
We assign to such open walks an algebraic area by closing them with a straight line from $(k,l)$
to the origin. It is clear than with this ``radial'' definition \cite{Desbois} the algebraic area measured in units of lattice
plaquettes can be half-integer.

We note that other closing prescriptions can be defined, such as, e.g., a ``rectangular'' prescription of
first returning vertically from then endpoint to the horizontal axis where the starting point lies and then
returning to the starting point horizontally. The algebraic areas for the two prescriptions are trivially related
by the area of the orthogonal triangle on the lattice with vertices on the two endpoints of the walk,
but they still group the walks differently in terms of their area and lead to different counting formulae.
We adopt the radial definition  as more natural and symmetric, although the rectangular one can also
be examined with minor modifications.

\subsection{Algebraic construction}

Similarly to closed walks, we define the algebraic area generating function of open walks as
\be
 G_{\bf {n}} (\Q) = \sum_A C_{\bf {n}} (A) \Q^{2A}
\label{Gn}\ee
with $C_{\bf {n}} (A)$ the number of walks of length ${\bf n}$ and area $A$ and $\Q$ a parameter dual to the area.
The exponent of $\Q$ was chosen to be $2A$ to avoid fractional powers arising from half-odd-integer
values of $A$.

The calculation of $G_{\bf {n}} (\Q)$ can be achieved by establishing an algebraic framework similar to the
one for closed walks, with an additional twist. Define operators $u,v,\sigma$ satisfying the defining relations
\be
v\,u=\Q^2 \, u\, v ~,~~ u\, \sigma = \sigma u^{-1} ~,~~ v\, \sigma = \sigma v^{-1} \label{uvs}
\ee
and a formal trace operation $\Tr (\,\cdot\,)$ on their algebra such that
\be
\Tr \sigma = \Tr (u\sigma) = \Tr (v \sigma) = 1~,~~ \Tr (v u \sigma) = \Q \label{Tr}
\ee
We define the Hamiltonian for the random walk as
\be
H = u+\ui+v+\vi
\ee
Then the area generating function is obtained as
\be
G_{\bf {n}} (\Q) = \Tr (H^{\bf {n}} \sigma)
\ee
The proof is along similar lines as in the closed walk case (where $\sigma$ is absent). Expanding $H^{\bf {n}}$
produces $4^{\bf {n}}$ monomials of the form $v^{l_i} u^{k_i} \cdots v^{l_1} u^{k_1}$, each corresponding to a
walk with $k_1$ horizontal steps followed by $l_1$ vertical steps etc., concluding with $k_i$ horizontal and $l_i$
vertical steps, and representing all possible walks with ${\bf {n}}$ steps.
Using $vu = \Q^2 uv$ to rearrange the terms brings $H^{\bf n}$ to the form
\be
H^{\bf {n}} = \sum_{k,l} g_{k,l} (\Q)\, v^l u^k ~,~~ |k\pm l| \le {\bf {n}}
\ee
reducing each walk ending at lattice point $(k,l)$ to a rectangular walk with $k$ horizontal steps followed by $l$ vertical steps and produces a coefficient $\Q^{2A'}$, with $A'$ the area between the original walk and the  corresponding  rectangular walk. $g_{k,l} (\Q)$ accounts for all paths ending on $(k,l)$ weighted by the corresponding area factors.
Finally, using commutation and trace relations (\ref{uvs},\ref{Tr}) we can show that
\be
\Tr (v^l u^k \sigma) = \Q^{kl}
\ee
$kl/2$ is the area of the rectangular walk closed with a straight line to the origin. Overall,
\be
\Tr (H^{\bf {n}} \sigma) = \sum_{k,l} g_{k,l} (\Q) \Tr (v^l u^k \sigma) = \sum_{k,l} g_{k,l} (\Q) \Q^{kl}
\ee
gives the full sum over walks of all possible endpoints weighted by $\Q^{2A}$, $A = A' +{kl \over 2}$ being
their total area, reproducing $G_{\bf {n}} (\Q)$.

\subsection{Representation of $u,v,\sigma$}

The main task is to evaluate the trace $\Tr (H^{\bf {n}} \sigma)$. This is made possible by finding an explicit matrix
representation for the operators $u,v,\sigma$, for which $\Tr$ would become the usual matrix trace $\tr$.

The algebra of $u,v$ is the standard clock-shift (or quantum torus) algebra and has finite dimensional
irreducible representations (irreps) for $\Q^2 = \exp(2i\pi p/q)$ with
$p,q$ mutually prime positive integers. The full $u,v,\sigma$ algebra (\ref{uvs}) has been analyzed
in \cite{hexagon} with the extra condition
$\sigma^2 =1$. Since $\sigma^2$ is a central element of the algebra, it becomes a constant in an
irrep and can be absorbed by an algebra-preserving redefinition $\sigma \to \lambda \sigma$,
so the irreps found in \cite{hexagon} also apply to our case.

In general, irreps are of size $2q$. In block form:
\be 
u=\begin{pmatrix}
u_o & 0\bb \\
0 & ~ u_o^{-1}\bb\\
\end{pmatrix}
,~~
v=\begin{pmatrix}
v_o & 0\bb \\
0 & ~v_o^{-1}\bb\\
\end{pmatrix}
,~~
\sigma=\begin{pmatrix}
0 & ~1\\
1 & ~0\\
\end{pmatrix},
\label{2qdim}\ee
with $u_o , v_o$ the $q$-dimensional irrep of the $u,v$ algebra. This representation, however, does not
fulfill the trace conditions (\ref{Tr}), giving vanishing traces. The remaining possibility is the reduced,
$q$-dimensional irrep that exists when the quantum torus Casimirs $u^q=e^{{\rm i}q k_x}$ and 
$v^q=e^{{\rm i} qk_y}$ become $\pm1$ ($k_x , k_y \in \{0,\pi / q\}$), 
and is given by the action on periodically
defined basis states $\ket{j}$ 
\bea
u\ket{j}=&{\rme}^{{\rmi}({k_x}+{2\pi p}j /q)}\ket{j} &,~~\ket{j}\equiv\ket{j \; (\text{mod}~q)} \nonumber\\ 
v\ket{j}=&\hskip -0.6cm {\rme}^{{\rmi}{k_y}}\ket{j\bb -\bb 1}&~~~~~~ 
\nonumber\\
\sigma \ket{j}=& {\rme}^{{\rmi}k_y (2j-r)} \ket{r\bb -\bb j} &,~~rp+{q k_x /\pi} = 0~ ({\text{mod}}~q) 
\label{qdim}.\eea
The ``pivot'' $r$ in the inversion action of $\sigma$ is $r=0$, if $k_x =0$, or the primary solution of the
Diophantine equation $kq-rp=1$, if $k_x =\pi /q$. Imposing the trace conditions (\ref{Tr}) further restricts
$q$ to an {\it odd} integer $q=2s+1$ (an even $q$ gives vanishing traces). 
It is convenient to fix the Casimirs $k_x = k_y =0$ and take $j$ in the range $-s \le j \le s$,
thus placing the pivot state $\ket{0}$ in the middle. We obtain the specific realization
\bea
&&u\ket{j}=\Q^{2j}\ket{j} ,~ v\ket{j}=\ket{j\bb -\bb 1},~
\sigma \ket{j}=\ket{-j} \label{ssigma}\\
&&\tr \sigma = \tr (u\sigma) = \tr (v\sigma) = 1 ~, ~~\tr(vu\sigma) = \Q \cr
{\rm with}&&
\Q= e^{i{2\pi p(s+1) \over 2s+1}}~, ~~ \ket{-s\bb-\bb1}\equiv\ket{s},~\ket{s\bb+\bb1}\equiv\ket{-s}\nonumber
\eea
Note that $\Q$ is a specific square root of the quantum torus algebra parameter
$\Q^2 = e^{i{2\pi p\over 2s+1}}$ and that $\Q^{2s+1} = 1$. This corresponds to
the $(2s\bb+\bb 1)$-dimensional matrix realization
{\small{\be
\hskip -0.5cm u = \begin{pmatrix}
\mbox{\smaller[2]{$\Q^{-2s}$}}& \bb0 & \cdots & 0 & \cdots & 0 & 0 \\
0 & \bb\mbox{\smaller[2]{$\Q^{-2s+2}$}} &\cdots  & 0& \cdots & 0 & 0 \\
\vdots & \bb\vdots &\bb\bb\bb \ddots & \vdots &\; \ddots & \vdots &\vdots\\
0 &\bb 0 &\bb\bb \cdots &1  & \cdots & 0 & 0 \\
\vdots &\bb \vdots & \bb\bb\ddots & \vdots & \;\ddots &\vdots & \vdots \\
0 &\bb 0 &\cdots & 0  & \cdots &\mbox{\smaller[2]{$\Q^{2s-2}$}} & 0 \\
0 &\bb 0 &\cdots & 0  & \cdots & 0 & \mbox{\smaller[2]{$\Q^{2s}$}}\\
\end{pmatrix}\bb,\,
\label{umat} v = \begin{pmatrix}
\;0\; & \;1\; & \;0\; & \;0\; & \cdots & \;0\; & \;0\; \\
0 & 0 & 1 & 0 &\cdots & 0 & 0 \\
0 & 0 & 0 & 1 &\cdots & 0 & 0\\
\vdots & \vdots & \vdots & \vdots &\ddots & \vdots & \vdots\\
0 & 0 & 0 & 0 &\cdots & 1 & 0\\
0 & 0 & 0 & 0 &\cdots & 0 & 1\\
1 & 0 & 0 & 0 & \cdots & 0 & 0 \\
\end{pmatrix}\bb,\,
\sigma = \begin{pmatrix}
0& \;\;0 & \cdots & 0 & \cdots & 0 & \;\; 1 \\
0 & \;\;0&\cdots  & 0& \cdots & 1 & \;\;0 \\
\vdots & \;\;\vdots & \ddots & \vdots &\; \iddots & \vdots &\;\;\vdots\\
0 & \;\; 0 & \cdots &1  & \cdots & 0 & \;\; 0 \\
\vdots &\;\; \vdots & \iddots & \vdots & \;\ddots &\vdots & \;\;\vdots \\
0 &\;\; 1 &\cdots & 0  & \cdots &0 & \;\;0 \\
1 & \;\; 0 &\cdots & 0  & \cdots & 0 &\;\; 0\\
\end{pmatrix}
\nonumber\ee}}

\subsection{Calculation of traces}

In the realization (\ref{ssigma}) $\sigma^2=1$, and the $(2s\bb+\bb 1)$-dimensional space decomposes into
an $(s\bb+\bb1)$-dimensional subspace with $\sigma = 1$ and an $s$-dimensional subspace with $\sigma = -1$.
The Hamiltonian $H = u + \ui + v + \vi$ commutes with $\sigma$, therefore
\be
\tr (H^{\bf {n}} \sigma) = \tr H_+^{\bf {n}} - \tr H_-^{\bf {n}} ~,~~~ H_\pm = H \small{1\pm \sigma \over 2}
\label{trpm}\ee
and we can evaluate the trace separately in each subspace.

To further facilitate the calculation, we use the trick \cite{kreft} of adopting a realization of $u,v,\sigma$
that eliminates the diagonal terms in $H$. This is achieved by the redefinition
\be
u \to \Q u v ~,~~ v \to v
\ee
One can check that the algebra and trace relations (\ref{uvs},\ref{Tr}) as well as
the Casimirs $u^q = v^q \he 1$ remain invariant under this redefinition. The Hamiltonian becomes
\be
H = \left(1+\Q u\right) v + \left(1+ \Q u^{-1} \right) v^{-1}
\label{newH}\ee
Choosing the basis $\ket{j}_\pm$ for the subspaces
\bea
&&\ket{j}_\pm = {1\over \sqrt 2} \bigl( \ket{j} \pm \ket{-j} \bigr) ~,~~ j \neq 0 \cr
&& \ket{0}_+ = \ket{0} ,~ \ket{0}_- = 0 ~;~~~ \ket{-j}_\pm = \pm \ket{j},~\ket{s+1}_\pm = \pm \ket{s}
\eea
the action of $H$ on these subspaces becomes
\bea
&&H \ket{j}_\pm = \left(1+\Q^{-2j-1} \right) \ket{j+1}_\pm + \sqrt{2}^{\,\delta_{j,1}}\left(1+\Q^{2j-1} \right) \ket{j-1}_\pm ~,
~~ j \neq 0,s \cr
&& H \ket{0}_+ = \sqrt{2} \left(1+\Q^{-1}\right) \ket{1}_+ \cr
&&H \ket{s}_\pm = \pm 2 \ket{s}_\pm + (1+\Q^{2s-1}) \ket{s-1}_\pm
\eea
We note that there is a single remaining diagonal term $\pm 2$ for $j=s$.

We can view both $H_+$ and $H_-$ as acting on the same states $\ket{j}$, $j=0,1,\dots,s$
with common matrix elements connecting states $j$ and $j\pm 1$, differing only on $0 \leftrightarrow 1$
and $s \to s$ transitions:
\be (H_+ )_{01} = (H_+ )_{10}^* = \sqrt 2 (1\hp \Q)~,~~
(H_- )_{01} = (H_- )_{10} = 0 ~,\;~~ (H_\pm )_{ss} = \pm 2
\ee
Traces can be expressed in terms of periodic ``paths'' of indices $i_1 , i_2 , \dots , i_{\bf {n}} , i_1$
\be
\tr H_\pm^{\bf {n}} = \sum_{i_1,i_2,\dots i_{\bf {n}}} (H_\pm )_{i_1 i_2} (H_\pm )_{i_2 i_3} \cdots (H_\pm )_{i_{\bf {n}} i_1}
\ee
The contribution of paths not going through $j=0$ and not containing $s \to s$ steps is the same for
$H_\pm$ and will cancel in (\ref{trpm}). Therefore, only paths that touch $0$ or ``creep" on $s$
will contribute. Further, all paths will have an equal number of steps $i\hm 1 \to i$ and $i \to i\hm 1$, and
each pair of such transitions will contribute an amplitude
\bea
s_i &=& \left(1+\Q^{-2i+1}\right) \left(1+\Q^{2i-1}\right) 
= \left( \Q^{-i+\half} + \Q^{i-\half} \right)^2 ~~i>1 \cr
s_1 &=& 2 \left(2 + \Q^{-1} + \Q \right) = 2 \left(\Q^{-\half} + \Q^\half \right)^2
\label{si}\eea

The above observations allow us to evaluate the trace in a combinatorial way, by examining separately
the cases of walks of even and odd length.

1) {\bf Even length ${\bf {n}}=2n$}: Paths must have an even number of steps $s \to s$, since all remaining steps come in pairs,
and these steps will contribute the {\it same} factor in $H_+^n$ and $H_-^n$. Therefore, such paths
will cancel unless they touch $0$. Only $H_+$ contributes for such paths. Assuming, for the moment,
$n\le s$, these paths cannot have any steps $s \to s$. (Paths of vertical increments $\pm 1$, are known as
Dyck paths in the mathematics literature.)
A typical such path is depicted in fig.~\ref{Near zero}.
{\begin{figure}
\centering 
{
\begin{tikzpicture}[scale=1]

\draw[help lines, gray] (0,-0.01) grid (10.5,6.01);

\tikzset{big arrow/.style={decoration={markings,mark=at position 1 with {\arrow[scale=3,#1,>=stealth]{>}}},postaction={decorate},},big arrow/.default=black}

\draw[very thick,-] (0,0) -- (11,0);
\draw[very thick,-] (0,-0.02) -- (0,7);
\draw[very thick,-] (0,6) node[left] {$s=6$} -- (10.5,6);

\draw[big arrow] (0,0) -- (11,0) node[below right] {$\bf n$};

\draw[big arrow] (0,-0.02) -- (0,7) node[left] {$i$};



%

\draw (0,3)node[left] {\color{black} $j=3$};
\draw[ultra thick,purple,-](0,1)node[left] {\color{black} $i_1=1$}--(1,2)--(2,1)--(3,0)--(4,1)--(5,2)--(6,3)--(8,1)--(9,2)--(10,1)node[above right] {\color{black} $i_{11} \hb=\hb i_1 \hb=\hb1$};

\fill (0,1) circle (2pt);
\fill (10,1) circle (2pt);
\fill (3,0) circle (2.2pt);

\end{tikzpicture}
}
\caption{\small{A typical path contributing to $\tr \hb (H^{\bf n} \,\hb\sigma )$ for $s\he 6$, with even length
${\bf n}\hb =\hb 2n \hb=\hb 10$, starting and ending at index $i \he 1$, and touching $0$
at $i_{4} =0$. It reaches a maximal level $j=3$ and has transitions $l_1 \he 1 , l_2 \he 3 , l_3 \he 1$.
}}\label{Near zero}
\end{figure}
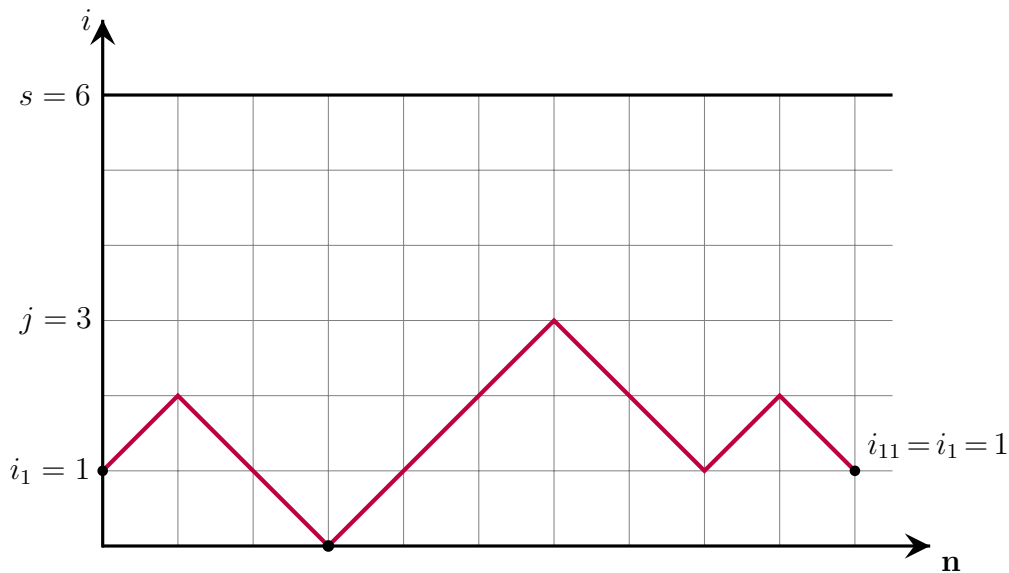}

Assuming the path reaches maximum level $j$, and calling $l_i$ ($1\le i \le j$) the number of (up or down)
transitions between levels $i\hm\hb 1$ and $i$, the trace is expressed as
\be
\tr (H^{2n} \sigma ) = \tr H_+^{2n} = \sum_{\sum_i l_i = n} 2n\, c(l_1, l_2,\dots,l_j )\, s_1^{l_1}
s_2^{l_2} \cdots s_j^{l_j}
\label{triven}\ee
The sum is over all nonvanishing integers summing to $n$, that is, over all compositions of $n$, and
$2n \, c(l_1,\dots,l_j )$ is the numbers of distinct periodic paths with the given number $l_i$ of transitions per level.
The counting of these paths is known, derived combinatorially or via a secular determinant method, and
takes the form\footnote{$c(l_1,\dots,l_j )$ is related to the $n^{th}$ cluster coefficient of identical
particles with quantum exclusion
statistics and was denoted $c_2 (l_1,\dots,l_j )$ in \cite{nous}, the index $2$ referring to exclusion of order 2.}
\be
c (l_1,l_2,\ldots,l_j) = {1 \over l_1} \prod_{i=1}^{j-1} {l_i + l_{i+1} -1 \choose l_{i+1}}
= {1 \over l_j} \prod_{i=1}^{j-1} {l_{i} + l_{i+1} -1 \choose l_{i}}
\label{civen}\ee
Combining (\ref{triven}), (\ref{si}) and (\ref{civen}), we obtain
\bea 
G_{2n} (\Q ) &=& 2n \bb\bb 
\sum_{{l_1, l_2, \ldots, l_j\atop \text{composition of}\, n}} \bb\bb 2^{l_1}\,  c(l_1, l_2,\dots,l_j )\,
\prod_{i=1}^j \Bigl( \Q^{-i+\half} + \Q^{i-\half} \Bigr)^{2l_i} \label{even} \\
&=&  2n \bb\bb\bb\sum_{{l_1, l_2, \ldots, l_j\atop \text{composition of}\, n}} \bb\bb\hb
{2^{l_1}\over \l_1} {(2\hp\Q^{-1}\hp\Q)^{l_1}} \prod_{i=2}^{j} {l_{i-1} + l_i -1 \choose l_i}
\hb \left(2 \hp \Q^{-2i+1} \hp \Q^{2i-1}\right)^{l_i} \nonumber
\eea

2) {\bf Odd length ${\bf {n}}=2n\hb-\hb 1$}:
Paths must have an odd number of steps $s \to s$, so these steps will contribute
{\it opposite} factors for $H_+$ and $H_-$. Assuming, again, $n\le s$, such paths never touch $0$ and thus
the total amplitude for $H_-^{2n-1}$ is the opposite of that for $H_+^{2n-1}$.
Such a path is depicted in fig.~\ref{Near s}.
{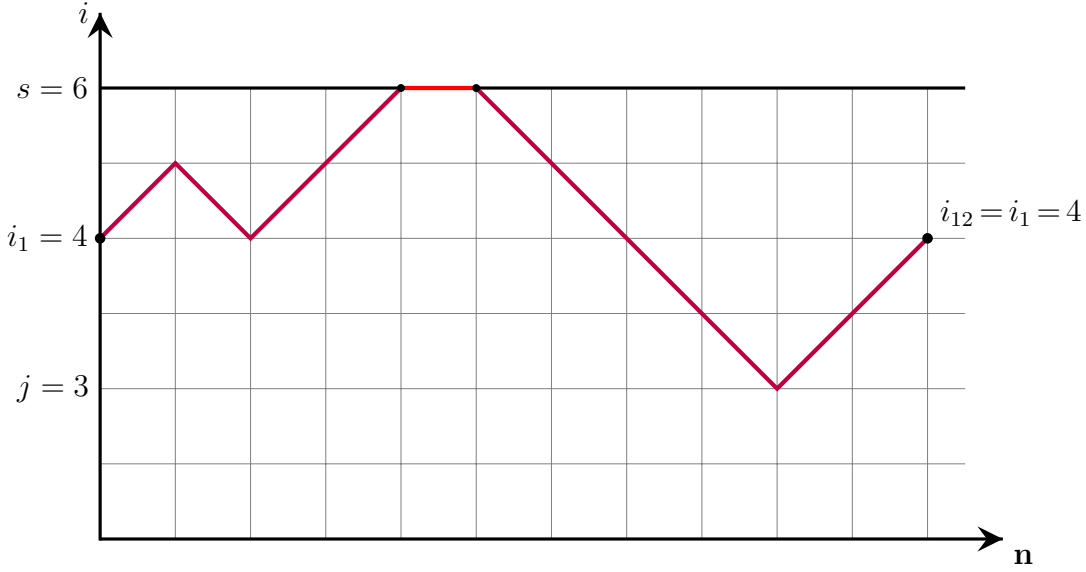
\begin{figure}
\centering 
{
\begin{tikzpicture}[scale=1]

\draw[help lines, gray] (0,-0.01) grid (11.5,6.01);

\tikzset{big arrow/.style={decoration={markings,mark=at position 1 with {\arrow[scale=3,#1,>=stealth]{>}}},postaction={decorate},},big arrow/.default=black}

\draw[very thick,-] (0,0) -- (12,0);
\draw[very thick,-] (0,-0.02) -- (0,7);
\draw[very thick,-] (0,6) node[left] {$s=6$} -- (11.5,6);

\draw[big arrow] (0,0) -- (12,0) node[below right] {$\bf n$};

\draw[big arrow] (0,-0.02) -- (0,7) node[left] {$i$};



%

\draw (0,2)node[left] {\color{black} $j=3$};
\draw[ultra thick,purple,-](0,4)node[left] {\color{black} $i_1=4$}--(1,5)--(2,4)--(4,6);
\draw[ultra thick,red,-](4,6)--(5,6);
\draw[ultra thick,purple,-](5,6)--(9,2)--(10,3)--(11,4)node[above right] {\color{black} $i_{12} \hb=\hb i_1 \hb=\hb 4$};

\fill (0,4) circle (2pt);
\fill (11,4) circle (2pt);
\fill (4,6) circle (1.5pt);\fill (5,6) circle (1.5pt);

\end{tikzpicture}
}
\caption{\small{A typical path contributing to $\tr \hb (H^{\bf n} \,\hb\sigma )$ for $s\he 6$, with odd length
${\bf n}\hb =\hb 2n+1 \hb=\hb 11$, starting and ending at index $i \he 4$, with one step on $i\he s$
between $i_5 \he i_6 \he 6$. It dips to a minimal level $s-j =2$ ($j\he 4$) and has transitions  
$l_0 \he 1, l_1 \he 1 , l_2 \he 2 , l_3 \he 1, l_4 \he 1$.
}}\label{Near s}
\end{figure}}

Assuming a path dips down to minimum level $s\hm j$, we call $l_i$ ($i \ge 1$) the number of (down or up)
transitions between level $s-i+1$ and $s-i$, and $2l_0 -1$ ($l_0 >0$) the (odd) number of $s \to s$ steps.
The total number of steps is $2l_0 \hb-\hb 1 + 2\sum_{i=1}^j l_i = 2n\hb-\hb1$, so $\sum_{i=0}^j l_i = n$.
The total trace can be expressed combinatorially as
\be
\tr(H^{2n-1} \sigma) = 2 \,\tr H_+^{2n-1} = 2 \sum_{\sum_i l_i = n}(2n\hm 1)\, {\bar c} (l_0,l_1,\dots, l_j )\,
2^{2l_0-1} s_s^{l_1} \cdots s_{s-j+1}^{l_j}
\ee
where $(2n\hm 1)\,{\bar c}(l_0, l_1 ,\dots, l_j )$ denotes the number of discrete periodic paths with the given number
of $s\to s$ steps and transitions. By taking each $s\to s$ step and extending it to an $s \to s\hp 1 \to s$ set
of transitions by adding a fictitious $s+1$ level, such paths become the mirror-image of paths touching
$0$ upon mapping levels $i \to s+1 -i$, so
\be
{\bar c} (l_0, l_1, \dots, l_j ) = c(2l_0\hm1,l_1,\dots,l_j) =
{1 \over 2l_0 \hm 1} {2l_0 + l_1 -2 \choose l_1} 
\prod_{i=1}^{j} {l_{i-1} + l_i -1 \choose l_i}
\label{barc}\ee
(The fact that the promotion of $s\to s$
to two transitions $s\to s\hp 1$ and $s\hp 1 \to s$ increases the length of the chain by $2l_0-1$ is
compensated by the fact that paths cannot start at $s\hp 1$.) Noting also that
\be
s_{s-i+1} = 2+ \Q^{-2s+2i-1} + \Q^{2s+2i+1} = 2 + \Q^{-2i} + \Q^{2i} = (\Q^{-i} + \Q^i )^2
\ee
we obtain the final result
\bea
\hskip -0.2cm G_{2n-1} (\Q ) &\bb=&\bb (2n\hm 1) \sum_{{l_0 , l_1, \ldots, l_j\atop \text{composition of}\, n}} \bb\bb{c} (2l_0\hm 1,l_1,\dots,l_j )
\prod_{i=0}^j (\Q^{-i} + \Q^i )^{2l_i} \label{odd}\\
&\bb=&\bb (2n\hm 1) \bb\bb 
\sum_{{l_0, l_1, \ldots, l_j\atop \text{composition of}\, n}} \bb\bb\bb
4^{l_0}{(2l_0\hp l_1\hm 2)!(l_0\hm 1)! \over (l_0\hp l_1\hm 1)! (2l_0\hm 1)!} \prod_{i=1}^{j} {l_{i-1} \hp l_i \hm 1 \choose l_i}\hb \left(\Q^{-i} \hp \Q^{i}\right)^{2l_i}\nonumber
\eea

\subsection{Generalization for all lengths and specific examples}\label{noumklapp}

We finally address the assumption made so far that ${\bf {n}}\hb\le\hb 2s$.
In general, for ${\bf {n}}\hb >\hb 2s$ states near both $\ket{0}$ and $\ket{s}$
need be considered and would lead to ``umklapp'' effects\footnote{By ``umklapp" effects we mean
walks with algebraic areas differing by multiples of $s\hp\half$ being counted together,
since $\Q^{2s+1}=1$. Their algebraic counteparts are paths of indices that wind around the periodic states
$\ket{i} = \ket{i\hp q}$, which in the $H_\pm$ formulation manifest as paths that both touch $0$ and
creep over $s$.}.
However, formulae (\ref{even}) and (\ref{odd})
do not involve $s$ explicitly, $\Q$ being the only parameter. Consequently, we can simply ignore the constraint
${\bf {n}}\le 2s$ and treat $\Q$ as a formal expansion parameter as in the original defining relation (\ref{Gn}).
Therefore, (\ref{even}, \ref{odd}) are valid for all values of ${\bf {n}}$ {\it without} restriction.

It is reassuring to give a few examples of the generating function formulae for low values of the length:
\bi
\item For length ${\bf {n}}=1$, setting $n=1$ in (\ref{odd}) only the term $l_0 =1$ survives and we obtain
$G_1 (\Q) = 4$.

\item For length ${\bf {n}}=2$, setting $n=1$ in (\ref{even}) only the $l_0=1$ term survives and we obtain
$G_2 (\Q) =8 + 4(\Q^{-1} + \Q)$

\item For length ${\bf {n}}=3$, setting $n=2$ in (\ref{odd}) only the $l_0 = 2$ and $l_0 = l_1 = 1$ terms
survive and we obtain $G_3 (\Q) 
= 40 + 12 (\Q^{-2} + \Q^2)$

\item For length ${\bf {n}}=4$, setting $n=2$ in (\ref{even}) only the $l_0 = 2$ and $l_0 = l_1 = 1$ terms
survive and we obtain $G_4 (\Q) = 80 + 48 (\Q^{-1} +\Q) + 16 (\Q^{-2} + \Q^2 )+ 16(\Q^{-3} + \Q^3)
+ 8(\Q^{-4} + \Q^4)$
\ei
It can be checked that these reproduce the correct number of open walks with the corresponding
length and area (exponent of $\Q^2$), and that the total number of walks, obtained by 
setting $\Q =1$ in the generating funcion $G_{\bf {n}}(\Q)$,  is $4^{\bf n}$ as required.

\subsection{Walk enumeration}

From the generating functions (\ref{even}) and (\ref{odd}) we can infer the number of paths
of given length and algebraic area $A/2$ by expanding in powers of $\Q$ and isolating the coefficient
of the term $\Q^A$. It is already clear from the form of (\ref{even}) that the expansion in powers of $\Q$\
will involve both even and odd powers, reflecting the fact that paths of even length can have half-integer
algebraic area, while (\ref{odd}) clearly involves only even powers, consistent with the fact that paths
of odd length can only have an integer area.

A binomial expansion of the upper expression in (\ref{even}) gives
\be
C_{2n} (A) = 2n \bb\bb\sum_{{l_1 , l_2, \ldots, l_j\atop \text{composition of}\, n}}
\bb\bb\bb 2^{l_1} \,{c} (l_1,\dots,l_j )\bb
\sum_{k_2=-l_2}^{l_2} \cdots \sum_{k_j = -l_j}^{l_j}
{2l_1 \choose l_1 \hp \sum_{r=2}^j (2r\hm 1)k_r \hm 2A }\prod_{s=2}^j {2l_s \choose l_s \hp k_s}
\label{Ceven}\ee
while a similar expansion of the upper expression in (\ref{odd}) gives
\be
C_{2n-1} (A) = (2n- 1) \bb\bb\sum_{{l_0 , l_1, \ldots, l_j\atop \text{composition of}\, n}}
\bb\bb\bb 4^{l_0}\, {c} (2l_0 -1,l_1,\dots,l_j )
\sum_{k_2=-l_2}^{l_2} \cdots \sum_{k_j = -l_j}^{l_j}
{2l_1 \choose l_1 \hp \sum_{r=2}^j r k_r \hm A }\prod_{s=2}^j {2l_s \choose l_s \hp k_s}
\label{Codd}\ee
In all expressions, binomial coefficients with entries outside of their range vanish and products with
lower term rank higher than the upper one become unity.

\subsection{Paths with fixed endpoints}\label{fixed}

We conclude with a brief discussion of the most general situation, namely, the enumeration of walks of
given length and area and with a fixed endpoint (the starting point is always placed at the origin).
As before, we will focus on evaluating the algebraic area generating function for such walks.

To fix the endpoint of the walks there are two possible approaches. One approach would be to consider
a generating function that assigns specific weights to the endpoint of the walk. This can be achieved by
considering operators $u,v$ with nontrivial Casimirs $u^q , v^q$ through the substitution
\be
u \to e^{i k_x} u ~,~~ v \to e^{i k_y} v
\ee
and evaluating $\tr (H^{\bf n} \sigma)$ as before. Note that, for such $u,v$,
\be
\sigma u = e^{2ik_x} u^{-1} \sigma~,~~
\sigma v = e^{2ik_y} v^{-1} \sigma~,~~
\tr (v^l u^k \sigma) = e^{ik k_x + il k_y} \Q^{kl}
\ee
So
\be
G_{\bf {n}} (\Q;k_x,k_y) = \tr (H^n \sigma)
\label{ktrace}\ee
reproduces the generating function of open walks, weighted by phase factors $ e^{-ik\, k_x -i l\, k_y}$ depending on their endpoint.
Walks ending at $k,l$ can then be isolated by
\be
{\tilde G}_{\bf {n}} (\Q;k,l) = {1\over 4\pi^2} \int_0^{2\pi} dk_x dk_y e^{-ik\, k_x -i l\, k_y}\, G_{\bf {n}} (\Q;k_x,k_y )
\label{tildeG}\ee
The explicit evaluation of the trace in (\ref{ktrace}), however, is nontrivial, as $H$ does not
commute with $\sigma$ any more.

An alternative approach would be to evaluate a general {\it matrix element} of the function $H^{\bf n}$, which,
as we shall explain, yields the generating function of walks ending on a straight line weighted
by their endpoint on this line. Specifically, consider
\be
G_{2n}^{(2I,J)} (\Q) := \bra{J\hm I} H^{2n} \ket{J\hp I} = \bigl<{|J\hm I|}\bigr| (H_+^{2n}
+ sgn(J^2 -I^2 ) H_-^{2n}) \bigl|{|J\hp I|}\bigr>
\label{ji}\ee
with $H$ as in (\ref{newH}), referring to the modified realization $u \to \Q u v$,
and states as defined in (\ref{ssigma}). Monomials $v^l u^k$ become
\be
v^l u^k \to v^l (\Q u v)^k = \Q^{-k^2} v^{l+k} u^k
\ee
and the $IJ$ matrix element becomes
\be
\bra{J\hp I} \Q^{-k^2} v^{l+k} u^k \ket{J\hm I} = \Q^{-k^2+2(J+I)k}\, \delta_{k+l,2I}
= Q^{2J k + lk} \,\delta_{l,2I-k}
\ee
The above gives the area-weighted sum of walks ending on even-parity points $k+l =2I$,
weighted by the factor $\Q^{2 J k}$ that depends on their final position $(k,2I-k)$. Multiplying by
$\Q^{-2 J k_0}$ and summing over $J$ would isolate the term
$k= k_0\,(\text{mod}(2s+1))$, thus reproducing the
generating function of walks ending at an even-parity sublattice point $(k_0,2I\hm k_0 )$ up to an
``umklapp'' periodicity $k_0 \sim k_0 + 2s\hp1$. The umklapp effect becomes relevant for walks
of length long enough to reach more than one periodic copies, and can be eliminated by assuming
\be
\sum_J \Q^{2 J x} = q\, \delta_{x,0}
\ee
thus ignoring the finiteness of $q=2s\hp 1$.
A similar construction generalizing (\ref{ji}) would work for walks ending on an odd-parity
sublattice point. The calculation of these matrix elements and corresponding sums is yet to be done.

\section{Conclusions and closed walks}

We conclude with some comments on closed walks of necessarily even length ${\bf n}=2n$,
for which an expression for their generating function and the corresponding algebraic area counting
formula are known \cite{bibi}. The methods in the present work offer an alternative way of calculating
these closed walk quantities, and one could hope to obtain alternative equivalent expressions.
The main reason for this hope is that the method based on traces involving $\sigma$, or corresponding
general matrix elements as in section \ref{fixed}, seems to at least partially evade ``umklapp'' effects,
as stressed in section \ref{noumklapp}.

In more detail, the first approach exposed here, based on $k_x , k_y$, has {\it no} umklapp effects
whatsoever. The second approach, relying on matrix elements for states $\ket{J\pm I}$,
has a {\it partial} umklapp effect relating to the position of the endpoint on the paradiagonal, but has
{\it no} umklapp effect with the position of the paradiagonal. The complexity of calculating
$G_{ {2n}}^{2I,J} (\Q)$, and especially $G_{ {2n}} (\Q; k_x , k_y )$, is the main impediment in
deriving expressions for closed walks, and their evaluations remains a task for the future.
 
Note that, putting $I=0$ in (\ref{ji}) and summing over $J$ amounts to calculating the trace
$\tr H^{2n}$, reproducing the known trace expression for the generating function of closed walks.
Similarly, the integral in (\ref{tildeG}) for $x=y=0$ would isolate terms
$v^0 u^0$ in $H^{2n}$ and would reduce $\tr (H^{2n} \sigma)$
to $(1/q) \tr H^{2n}$, again reproducing the known closed walk counting formula in terms of the trace.
It is the evaluation
of $G_{ {2n}} (\Q;k_x,k_y)$ as an explicit function of $k_x,k_y$, using the techniques of the present work,
that might yield alternative formulae. This is yet to be accomplished.

\vskip 0.5cm

\noindent
{\bf Acknowledgement}

\noindent
The work of A.P. was supported in part by NSF under grant NSF-PHY-2112729 and by a PSC-CUNY grant.

\appendix
\section*{Appendices}
\addcontentsline{toc}{section}{Appendices}
\renewcommand{\thesubsection}{\Alph{subsection}}

\subsection{``Diagonal'' walks}

The simplest case is walks that start at the origin and end anywhere along the line of lattice
points with coordinates $k+l =0$. Such walks necessarily have an even length.
The evaluation of their generating function $G^{(0)}$ is achieved by simply calculating
the $00$ matrix element
\be
G_{2n}^{(0)} (\Q)= \bra{0} H^{2n} \ket{0}
\ee
with $H$ as in (\ref{newH}), referring to the modified representation $u \to \Q u v$,
and states as defined in (\ref{ssigma}). Monomials $v^l u^k$ become in this representation
\be
v^l u^k \to v^l (\Q u v)^k = \Q^{-k^2} v^{l+k} u^k
\ee
and the $00$ matrix element becomes
\be
\bra{0} \Q^{-k^2} v^{l+k} u^k \ket{0} = \Q^{-k^2} \delta_{k+l}
\ee
constraining the walks to the $k+l=0$ subset and reproducing the radial area of such walks. We note that
walks ending on the diagonal $k=l$ trivially have the same generating function, due to the invariance of
the algebraic area under lattice $\pi/2$ rotations.

The evaluation of the matrix element is done in analogy to the trace of $H^n \sigma$: since
$\sigma \ket{0} = \ket{0}$, only the part $\bra{0} H_+^{2n} \ket{0}$ will contribute. We obtain
an expression similar to (\ref{triven}), with the difference that now the combinatorial prefactor
counts only paths that start from $0$ and end at $0$. This counting differs from the
full counting $2n c(l_0,\dots,l_j)$ by a factor of $l_1 /(2n)$. We obtain
\bea 
G_{2n}^{(0)} (\Q ) &=&  
\sum_{{l_1, l_2, \ldots, l_j\atop \text{composition of}\, n}} \bb\bb 2^{l_1} \, l_1 \,  c(l_1, l_2,\dots,l_j )\,\prod_{i=1}^j \Bigl(
\Q^{-i+\half} + \Q^{i-\half} \Bigr)^{2l_i} \label{0even} \\
&=& \bb\bb\sum_{{l_1, l_2, \ldots, l_j\atop \text{composition of}\, n}} \bb\bb
{2^{l_1}} {(2\hp\Q^{-1}\hp\Q)^{l_1}} \prod_{i=2}^{j} {l_{i-1} + l_i -1 \choose l_i}\hb \left(2 \hp \Q^{-2i+1} \hp \Q^{2i-1}\right)^{l_i} \nonumber
\eea
and a corresponding expression for the enumeration of walks
\be
C_{2n}^{(0)} (A) = \bb\bb\sum_{{l_1 , l_2, \ldots, l_j\atop \text{composition of}\, n}}
\bb\bb\bb 2^{l_1} \bb
\sum_{k_2=-l_2}^{l_2}\bb \cdots\bb \sum_{k_j = -l_j}^{l_j}
{2l_1 \choose l_1 \hp \sum_{r=2}^j (2r\hm 1)k_r \hm 2A }\prod_{i=2}^{j} {l_{i-1} \hp l_i \hm 1 \choose l_i}
{2l_i \choose l_i \hp k_i}
\label{C0}\ee

\subsection{``Paradiagonal'' walks}

With a similar reasoning, we can consider walks that end in an even-parity paradiagonal $k+l = 2 I$, $I \ge 0$.
(The cases $I<0$, and $k-l=2I$, are, again, trivially related to the present one by $\pi$ or $\pi/2$
lattice rotations.) These have as generating function
\be
G_{2n}^{(2I)} (\Q) = \bra{-I} H^{2n} \ket{I} = \bra{I} H_+^{2n} \ket{I} - \bra{I} H_-^{2n} \ket{I}
\ee
Only index paths that touch $0$ contribute, through $\bra{I} H_+^{2n} \ket{I}$.

The evaluation of
the matrix element procees similarly to the case $I=0$: we call again $l_i$
($1 \le i \le j$) the number of up or down transitions between levels $i\hm 1$ and $i$, where now necessarily
$j \ge I$. The number of paths $P(I;l_1,\dots,l_j)$ starting and ending at level $I$ with steps $l_1,\dots,l_j$
is given by
\be
P(I;l_1 , \dots , l_j ) = \left(l_I + l_{I+1} \right) c(l_1, l_2, \dots, l_j ) ~~~(l_0 = l_{j+1} \equiv 0)
\ee
with $c(l_1,\dots,l_j)$ as in (\ref{civen}). (The term proportional to $l_I$ above is the number of paths
starting downwards at $I$, and the term proportional to $l_{I+1}$ is the number of paths starting
upwards.) Note that summing over all $I$ gives
\bea
\sum_{I=0}^j P(I;l_1,\dots,l_j) &=& \left[l_1 + (l_1\hp l_1) + \cdots + (l_{j-1} \hp l_j ) + l_j\right]
 c(l_1, \ldots, l_j ) \cr
&=& 2 (l_1 + \cdots + l_j )  \, c(l_1, \dots, l_j ) = 2n \, c(l_1, \dots, l_j )
\eea
reproducing the full counting of paths for unrestricted walks. Overall we obtain
\be
G_{2n}^{(2I)} (\Q ) =
\sum_{{l_1, l_2, \ldots, l_j ;\, j\ge I\atop \text{composition of}\, n}} \bb\bb 2^{l_1} \, (l_I + l_{I+1}) \,  c(l_1, l_2,\dots,l_j )\,\prod_{i=1}^j \Bigl(
\Q^{-i+\half} + \Q^{i-\half} \Bigr)^{2l_i}
\label{Ieven}
\ee
Clearly (\ref{0even}) is a special case of (\ref{Ieven}) with $I=0$.

Odd-parity paradiagonal walks ending at $x+y=2I\hp 1$ ($I\ge 0$) necessarily have an odd length $2n-1$.
Their generating function can be expressed as
\be
G_{2n-1}^{(2I-1)} = \bra{s\hm I} H^{2n-1} \ket{-s\hp I} = \bra{s\hm I} (H_+^{2n-1} - H_-^{2n-1}) 
\ket{s\hm I}
\label{oddparity}\ee
Index paths for such walks will have an odd number of transitions $s \to s$, so the contributions of
$H_+$ and $H_-$ are equal. Assuming an index path that dips down to minimum level $s-j$, where necessarily
$j\ge I$, we call $l_i$ ($i \ge 1$)
the number of (down or up) transitions between level $s-i+1$ and $s-i$, and $2l_0 -1$ ($l_0 >0$) the
(odd) number of $s \to s$ steps. Using the trick of adding a fictitious $s\hp 1$ level and mapping
levels $i \to s\hp 1\hm i$, the number of index paths that start and end at level $s\hm I$ becomes
\bea
{\bar c}(I;l_0,l_1,\dots,l_j) &=& (l_I + l_{I+1} )\, c(2l_0\hm1,l_1,\dots,l_j ) ~~,~~~~I\ge 1 \\
&=& (2l_0 \hm 1 + l_1 )\, c(2l_0\hm1, l_1,\dots,l_j ) ~,~~ I=0 \nonumber
\eea
Note that summing over all $I$ we have
\be
\sum_{I=0} {\bar c}(I;l_0,l_1,\dots,l_j) = 2l_0 \hm 1 + 2l_1 + \cdots + 2l_j = 2n -1
\ee
reproducing the full counting $(2n\hm 1)\, {\bar c} (l_0,\dots,l_j )$ of (\ref{barc}). Overall we obtain
\be
G_{2n}^{(2I+1)} (\Q ) =
\sum_{{l_0, l_1, \ldots, l_j ;\, j\ge I\atop \text{composition of}\, n}} \bb\bb (l_I + l_{I+1}) \,  c(2l_0\hm 1,l_1, \dots,l_j )\,\prod_{i=0}^j \Bigl(
\Q^{-i} + \Q^i \Bigr)^{2l_i}
\label{Iodd}
\ee

The number of paradiagonal walks of fixed length and area can be found by isolating the term
$\Q^{2A}$ in expressions (\ref{Ieven},\ref{Iodd}), as in (\ref{Ceven},\ref{Codd},\ref{C0})
and we will not write the explicit formulae.

\end{document}